\documentstyle[PASJadd]{PASJ95}
\input epsf
\begin{document}

\title{Detecting Galactic MACHOs with VERA through Astrometric Microlensing of Distant Radio Sources}
\author{Mareki {\sc Honma}$^{1,2}$\\
$^1${\it VERA Project Office, National Astronomical Observatory of Japan, Mitaka 181-8588, Japan}\\
$^2${\it Earth Rotation Division, National Astronomical Observatory of Japan, Mizusawa 023-0861, Japan}\\
{\it E-mail: honmamr@cc.nao.ac.jp}}

\abst{
In this paper we investigate the properties of astrometric microlensing of distant radio sources (QSOs and radio galaxies) due to MACHOs, and discuss their implications for VERA (VLBI Exploration of Radio Astrometry).
First we show that in case of astrometric microlensing of distant sources, the event duration is only a function of the lens mass and tangential velocity, but independent of the lens distance, in contrast to the well-known three-fold degeneracy for photometric microlensing.
Moreover, the lens mass $M$ is scaled by the tangential velocity $v_\perp$ as $M\propto v_\perp$, rather than $M\propto v_\perp^2$ which is the case for photometric microlensing.
Thus, in astrometric microlensing the dependence of the lens mass on the unknown parameter $v_\perp$ is weaker, indicating that the duration of astrometric microlensing event is a better quantity to study the mass of lensing objects than that of photometric microlensing.
We also calculate the optical depth and event rate, and show that within 20$^\circ$ of the galactic center a typical event rate for 10 $\mu$as-level shift is larger than $2.5 \times 10^{-4}$ event per year, assuming that a quarter of the halo is made up with MACHOs.
This indicates that if one monitors a few hundred sources for $\sim$20 years, such an astrometric microlensing event is detectable.
Since a typical event duration is found to be fairly long (5 to 15 years), the frequency of the monitoring observation can be relatively low, i.e., once per six months, which is rather reasonable for practical observations.
We discuss practical strategy for observing astrometric microlensing with VERA, and argue that an astrometric microlensing event due to MACHOs can be detected by VERA within a few decades.
}
\kword{Astrometry --- The Galaxy: halo --- Gravitational Microlensing --- Radio Continuum: Galaxies --- VERA}
\maketitle
\thispagestyle{headings}

%%%
\section{introduction}
%%%
Gravitational microlensing is one of the most powerful tools to trace the Massive Astrophysical Compact Halo Objects (MACHOs) in the Galaxy's halo.
After the first proposal by Paczynski(1986), a number of extensive searches for gravitational microlensing have been performed toward the Magellanic clouds and the Galactic bulge.
In such studies, more than a million of stars are being monitored frequently, typically once per night, to look for the brightness variation caused by gravitational microlensing.
Up to date, searches for such 'photometric microlensing' events have detected more than a dozen of events toward the Magellanic Clouds (e.g., Alcock et al. 1993; 1997; 2000; Aubourg et a. 1993).
Based on a statistical study of the detected events, the MACHO collaboration (Alcock et al. 1997; 2000) have obtained the lens mass of $0.5 M_\odot$, and suggested that the MACHOs in the Galactic halo may be old white dwarfs.
However, the lens mass is strongly dependent on the halo model, and a more or less massive lens is also consistent with the observations depending on the halo model of the Galaxy (e.g., Honma \& Kan-ya 1998).
Moreover, the uncertainty in lens distance allows different interpretations of observed microlensing events other than MACHOs.
In fact, there are several arguments that the lensing objects may be stars in warped disk (e.g., Evans et al. 1998), stars in previously-unknown intervening objects (e.g., Zaritsky \& Lin 1997; Zhao 1998), or stars in the Magellanic clouds themselves (e.g., Sahu 1994).
Thus, the nature of MACHOs remains unclear.

In order to reveal the nature of MACHOs, some additional information other than what comes from photometric lensing is highly important.
One potential way to extract additional information for microlensing events is to measure a positional shift due to microlensing.
Several investigations have been made to discuss the implications of such an astrometric observation of microlensing events, demonstrating that such a positional shift can be detected by upcoming space satellite missions, and also that from astrometric observation one can determine the lens parameters such as lens distance and lens mass (e.g., Hog et al. 1995; Miyamoto \& Yoshii 1995; Walker 1995).
These astrometric observations are to be performed after the event detection based on the photometric monitoring, and in this sense, supplementary to photometric microlensing event.

On the other hand, it is also known that astrometric observation alone can be used to study the nature of lensing objects.
For instance, it has been argued that an astrometric effect of gravitational microlensing can be used to measure the mass of a nearby stars (e.g., Hosokawa et al. 1993; Paczynski 1996; 1998; Miralda-Escude 1996).
Recently, the fluctuation of the radio reference frame by microlensing due to disk stars and MACHOs in the Galaxy has been studied by Hosokawa et al.(1997), and possibility to observe astrometric microlensing events with ground-based or space optical-interferometer have been discussed by Miralda-Escude (1996), Boden et al.(1998), Paczynski (1998), and Dominik \& Sahu (2000). 
These studies revealed that astrometric microlensing events are practically detectable and useful for studying the nature of lensing objects, if we obtain a position accuracy of 10 $\mu$as level.
Thus, the astrometric microlensing will soon become one of major tools to study lensing objects.

Following the recent studies on astrometric microlensing, in this paper we investigate astrometric microlensing of distant sources (QSOs and radio sources) due to MACHOs, focusing on the possible application to the VERA project (VLBI Exploration of Radio Astrometry).
The VERA project (see for detail Sasao 1996; Kawaguchi et al. 2000; Honma et al.2000), which aims at radio astrometry at 10 $\mu$as level, has been approved in 2000, and is anticipated to start its observation by 2003.
Distant radio sources like QSOs and radio galaxies, which are one of major observational objects of VERA, are potential targets for tracing astrometric microlensing with VERA, but few studies have been made concerning the astrometric microlensing of distant radio sources after the pioneering work by Hosokawa et al.(1997).
For these reasons, we study the astrometric microlensing of distant radio sources and discuss the implications for VERA.

The plan of this paper is as follows; in section 2, we briefly review on astrometric microlensing, in particular parameters that can be extracted from the image trajectory.
In section 3 we investigate the event duration of astrometric microlensing.
In section 4 and 5, we calculate optical depth and event rate, and show that one can detect an astrometric microlensing based on monitoring of a few hundred sources for 10 to 20 years.
Finally in section 6, we discuss the implication of astrometric microlensing for VERA, and present a practical strategy for detecting astrometric microlensing.

%%%
\section{Brief Review on Astrometric Microlensing}
%%%
\subsection{Astrometric Microlensing Event}

In this chapter we briefly review the characteristics of astrometric microlensing.
When a source is closely aligned with a lens, the gravitational deflection of light produces two images whose positions are given by (e.g., Paczynski 1986)
\begin{equation}
\label{eq:r1,2}
r_{1,2}=\frac{u\pm\sqrt{u^2+4}}{2}.
\end{equation}
Here $r_{1,2}$ is the separation between the lens and the two images in the lens plane, and $u$ is the separation between the source and the lens in the lens plane.
Note that both $r_{1,2}$ and $u$ are normalized with the Einstein ring radius $R_{\rm E}$; for instance, the lens-source separation $u$ can be written as
\begin{equation}
u=\frac{R}{R_{\rm E}},
\end{equation}
where $R$ is the lens-source distance projected onto the lens plane.
As widely-known, the Einstein radius $R_{\rm E}$ is given by (e.g., Paczynski 1986)
\begin{equation}
\label{eq:E-ring}
R_{\rm E} =\sqrt{\frac{4GM}{c^2}\frac{D_{\rm d} D_{\rm ds}}{D_{\rm s}}}.
\end{equation}
Here $M$ is the lens mass, and $D_{\rm d}$, $D_{\rm s}$, and $D_{\rm ds}$, are lens distance, source distance and lens-source distance, respectively.

Here we introduce a coordinate system on the lens plane where the source is at the origin and the lens is moving parallel to the $x$ axis (figure 1).
The origin of time $t=0$ is set at the time of the closest approach between the lens and the source.
In such a coordinate system, the lens position is described as $(v_\perp t,\; -\beta R_{\rm E})$, where $v_\perp$ is the tangential velocity of the lens, and $\beta$ is the impact parameter normalized with $R_{\rm E}$ (note that $\beta$ is always positive).
With the parameters described above, the lens-source distance $R$ can be written as
\begin{equation}
R=\sqrt{(v_\perp t)^2 + (\beta R_{\rm E})^2}.
\end{equation}

Here we calculate the positional shift of sources due to gravitational microlensing.
Generally, two images produced by gravitational microlensing cannot be resolved, but the centroid of two images can be observed to trace the astrometric effect in microlensing.
If the lens is far from the source in the lens plane (i.e., $u\gg 1$), the secondary image becomes too faint, and the image centroid coincides with that of the primary image.
Since most of astrometric microlensing events with 10 $\mu$as-level positional shift have $u\gg 1$ (as we will see in next sections), in the present paper we concentrate on the case of $u\gg 1$, and approximate the image centroid motion by that of the primary image (for more general treatment, see Boden et al.1998; Dominik \& Sahu 2000).

\begin{figure}[t]
\vspace{6cm}
\epsfxsize=120pt
\epsfbox[130 330 300 500]{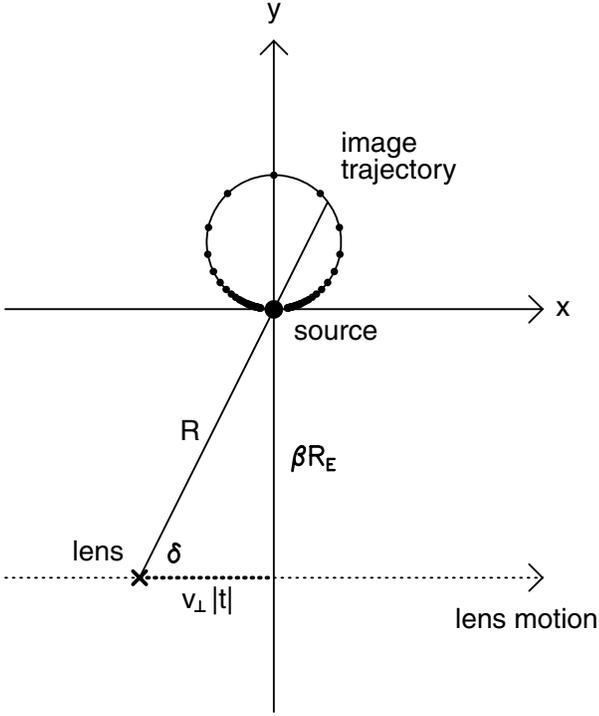}
\caption{Schematic view of the location of lens, source, and lensed image in the lens plane. Trajectory of an image is a circle in case of $u\gg 1$.
Dots along the trajectory circle show the image position with a constant time interval, $0.4 t_{\rm q}$.
}
\end{figure}

In case that $u\gg 1$, the position of primary image (eq.[\ref{eq:r1,2}]) is well approximated by,
\begin{equation}
r = u + \frac{1}{u}.
\end{equation}
The first term $u$ corresponds to the lens-source separation in the lens plane, and the second term $1/u$ describes the image shift due to gravitational microlensing.
Hence, the angular positional shift of the image relative to the original source position is obtained as
\begin{equation}
\theta_{\rm s} = \frac{1}{u}\; \theta_{\rm E}.
\end{equation}
Here $\theta_{\rm E}$ is the angular Einstein radius which is related to the Einstein ring radius $R_{\rm E}$ as
\begin{equation}
\theta_{\rm E} = \frac{R_{\rm E}}{D_{\rm d}}.
\end{equation}
The angular position of the image in the lens plane can be described as 
\begin{equation}
(\theta_x,\theta_y) = (\theta_{\rm s}\cos \delta,\; \theta_{\rm s}\sin \delta),
\end{equation}
where the angle $\delta$ is defined as (see figure 1)
\begin{equation}
\tan \delta = -\frac{\beta R_{\rm E}}{v_\perp t}.
\end{equation}

Using equations described above, one may rewrite the image position as
\begin{equation}
(\theta_x,\; \theta_y) = \left(\frac{-(\theta_{\rm E}/\beta)(t/\beta t_{\rm E})}{(t/\beta t_{\rm E})^2 + 1},\; \frac{\theta_{\rm E}/\beta}{(t/\beta t_{\rm E})^2 + 1}\right),
\end{equation}
where we have introduced the Einstein ring crossing time $t_{\rm E}$ which is defined as
\begin{equation}
t_{\rm E} \equiv \frac{R_{\rm E}}{v_\perp}.
\end{equation}

After a simple calculation, one may easily find the following relation between $\theta_x$ and $\theta_y$,
\begin{equation}
\theta_x^2 + \left(\theta_y-\frac{\theta_{\rm E}}{2\beta}\right)^2 = \left(\frac{\theta_{\rm E}}{2\beta}\right)^2.
\end{equation}
Hence, the trajectory of the image is a circle centered on $(0,\; \theta_{\rm E}/2\beta )$ with a diameter of $\theta_{\rm E}/\beta$ (Boden et al.1998; Dominik \& Sahu 2000).
We hereafter refer to this image trajectory as {`}trajectory circle{'}.
In figure 1, we show a schematic view of the image motion along the trajectory circle.
The time intervals between adjacent dots are equal to 0.4$\times (\beta t_{\rm E})$.
As seen in figure 1, the image motion is faster in the upper half of the trajectory circle than in the lower half, and vanishes toward the origin where the source should be located in case of no microlensing.
The largest image shift parallel to the x-axis occurs when $t_{\rm q}=\pm \beta t_{\rm E}$, i.e., $R =\sqrt{2}\beta R_{\rm E}$.
Note that an image goes through an upper quarter of the trajectory circle within the time interval of $t_{\rm q}$.
Thus, $t_{\rm q}$ gives a characteristic time-scale of an astrometric microlensing event.
Note that from the observation of trajectory circle, one can obtain $t_{\rm q}(=\beta t_{\rm E})$ and $\theta_{\rm E}/\beta$.

\subsection{Angular Size of Astrometric Lens}

Here we estimate the angular size of astrometric lenses.
Since we mainly consider the implications of astrometric microlensing for VERA, in the following analysis we assume that the sources are radio galaxies and QSOs which are much more distant (at least a few Mpc) than the lens (typically $\sim$10 kpc).
In such a case, the source distance $D_{\rm s}$ and the lens-source distance $D_{\rm ds}$ are much greater than the lens distance $D_{\rm d}$, and the Einstein ring radius (equation [\ref{eq:E-ring}]) is well approximated as
\begin{equation}
\label{eq:E-ring-approx}
R_{\rm E} =\sqrt{\frac{4GM}{c^2}{D_{\rm d}}}.
\end{equation}
Similarly, the angular Einstein ring radius $\theta_{\rm E}$ may be approximated as
\begin{equation}
\label{eq:angular-E-ring-approx}
\theta_{\rm E} = \sqrt{\frac{4GM}{c^2 D_{\rm d}}}.
\end{equation}
If one substitutes typical parameters of MACHOs (e.g., Alcock et al. 1997; 2000), the angular Einstein ring can be estimated as
\begin{equation}
\theta_{\rm E} = 0.72  \times \left(\frac{M}{0.5 M_\odot}\right)^{1/2} \left(\frac{D_{\rm d}}{8 {\rm kpc}}\right)^{-1/2}\;\;\; {\rm mas}.
\end{equation}
Since the angular diameter of lens trajectory is given by $\theta_{\rm E}/\beta$, the maximum value of $\beta$ for a detectable event is given by 
\begin{equation}
\label{eq:beta_max}
\beta_{\rm max} = \theta_{\rm E}/\theta_{\rm min},
\end{equation}
where $\theta_{\rm min}$ is the minimum position shift that can be measured.
The astrometric projects planned to start within a decade (e.g., {\it SIM}, {\it GAIA}, {\it VERA}) aims at the position accuracy of 10 $\mu$as or higher.
Since we consider the observation of astrometric microlensing with those projects (in particular VERA), we set $\theta_{\min}=10$ $\mu$as.
Thus, the maximum value of $\beta$ ($\beta_{\rm max}$) is given as follows,
\begin{eqnarray}
\beta_{\rm max} &=& 72 \times \left(\frac{M}{0.5 M_\odot}\right)^{1/2} \nonumber\\
 & & \qquad \times \left(\frac{D_{\rm d}}{8 {\rm kpc}}\right)^{-1/2} \left(\frac{\theta_{\rm min}}{10 \mu{\rm as}}\right)^{-1}.
\end{eqnarray}
This indicates that the astrometric microlensing is detectable even when the minimum lens-source separation is larger than the Einstein ring radius by several tens.
This makes the probability of astrometric microlensing much higher than the normal microlensing.

\begin{figure}
\vspace{7cm}
\epsfxsize=35pt
\epsfbox[220 130 270 180]{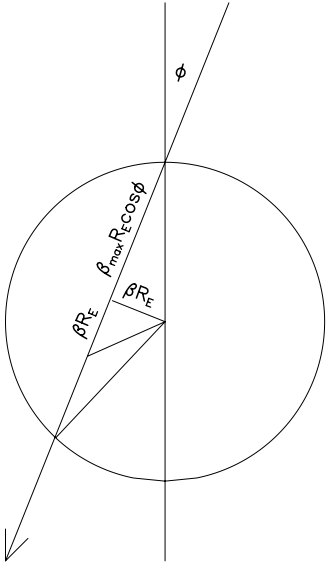}
\caption{Schematic view of the motion of source relative to lens. The large circle gives the lens size, which is equal to $\beta_{\rm max}R_{\rm E}$.
}
\end{figure}

%%%
\section{Event Duration}
%%%
\subsection{Basic Equations}

In this section, we investigate the event duration of astrometric microlensing.
Usually, the duration of microlensing event is defined as the time during which the source is inside the lens.
For the photometric microlensing, the lens size is equal to the Einstein ring radius $R_{\rm E}$, and the event duration $t_{\rm ph}$ is given as the time for which $R\le R_{\rm E}$.
Similarly to the photometric microlensing case, the lens size of an astrometric microlensing event is given by $\beta_{\rm max} R_{\rm E}$, and the duration of an astrometric microlensing event is defined as the time for which $R\le \beta_{\rm max} R_{\rm  E}$ (in other words, the time during which the image position shift is larger than the accuracy of position measurement $\theta_{\rm min}$).
Thus, the event duration $t_{\rm ast}$ can be written as
\begin{equation}
\label{eq:t_ast}
t_{\rm ast} = \frac{2\beta_{\rm max} R_{\rm E}}{v_\perp} \cos \phi.
\end{equation}
See figure 2 for a schematic view of lens-source geometry and the definition of the angle $\phi$.
The average event duration can be obtained by integrating $t_{\rm ast}$ over possible range of $\phi$.
Since the probability of an event with a certain value of $\phi$ is proportional to $\cos \phi$, the average event duration can be given as follows,
\begin{equation}
\label{eq:t_ast_av_def}
\bar{t}_{\rm ast}=\frac{\int t_{\rm ast} \cos \phi\; d\phi}{\int \cos \phi\; d\phi}.
\end{equation}
Substituting equation (\ref{eq:t_ast}) into equation (\ref{eq:t_ast_av_def}) and performing the integration with respect to $\phi$, one obtains that 
\begin{equation}
\label{eq:t_ast_av_with_beta}
\bar{t}_{\rm ast} = \frac{\pi}{2} \left(\frac{\beta_{\rm max}R_{\rm E}}{v_\perp}\right).
\end{equation}
Substituting the expressions of $R_{\rm E}$ and $\beta_{\rm max}$ (equations [\ref{eq:E-ring-approx}], [\ref{eq:angular-E-ring-approx}], and [\ref{eq:beta_max}]), one can obtain
\begin{equation}
\label{eq:t_ast_av}
\bar{t}_{\rm ast} = \frac{2\pi G M}{c^2 v_\perp \theta_{\rm min}}.
\end{equation}
Interestingly, the average duration $\bar{t}_{\rm ast}$ contains only two unknown parameters, the lens mass $M$ and its tangential velocity $v_\perp$, but independent of the lens distance $D_{\rm d}$.
Hence, the degree of degeneracy in the event duration is lower than that in case of photometric microlensing, where the event duration depends on three parameter; $v_\perp$, $D_{\rm d}$ and $M$ (note that this independence of the lens distance comes from the assumption of distant source. If this assumption is invalid, the event duration $\bar{t}_{\rm ast}$ depends also on the lens distance $D_{\rm d}$).
Moreover, the average event duration $\bar{t}_{\rm ast}$ is proportional to the lens mass $M$, and inversely proportional to the tangential velocity $v_\perp$.
Thus, once the average event duration is observed, the mass is scaled as $M\propto v_\perp \bar{t}_{\rm ast}$.
This dependence contrasts to that of photometric microlensing, where the event duration $\bar{t}_{\rm ph}$ is proportional to $M^{1/2} v_\perp^{-1}$ (e.g., Paczynski 1986).
Hence, for photometric microlensing events, the lens mass $M$ is scaled as $M\propto (v_\perp \bar{t}_{\rm ph})^2$, depending more strongly on the unknown parameter $v_\perp$ than in case of astrometric microlensing.
This indicates that the event duration of astrometric microlensing is a better quantity to investigate the mass and the nature of MACHOs rather than that of photometric microlensing.

\subsection{Halo Microlensing}

Here we estimate a typical value of the event duration for halo MACHO case based on the same parameters used in the previous sections.
Substituting $M=0.5M_\odot$, $v_\perp=200$ km s$^{-1}$, and $\theta_{\rm min}=10$ $\mu$as, the event duration is obtained as
\begin{eqnarray}
\bar{t}_{\rm ast} &=& 15.3 \times \left(\frac{M}{0.5 M_\odot}\right) \nonumber \\
& & \times \left(\frac{v_\perp}{200 {\rm km\; s}^{-1}}\right)^{-1} \left(\frac{\theta_{\rm min}}{10 \mu{\rm as}}\right)^{-1}\; {\rm yr}.
\end{eqnarray}
As seen in above equation, the event duration of astrometric microlensing event is significantly longer than that of photometric microlensing event.
This is due to large $\beta_{\rm max}$ for astrometric microlensing, typically $\sim$70, while $\beta_{\rm max}$ is unity for photometric microlensing.
In fact, if one takes a typical value of $t_{\rm E}=40$ days from LMC microlensing (e.g. Alcock et al. 1997; 2000), a rough estimate gives that $t_{\rm ast}\sim 2 \beta_{\rm max} t_{\rm E} = 15$ yr.

An event duration of $\sim$ 15 yr may seem to be too long for practical monitoring observation.
This is because the event duration $t_{\rm ast}$ is the time during which the image shift is larger than $\theta_{\rm min}$.
On the other hand, we have already shown in figure 1 that the image trajectory is a circle, and that the image motion is faster in the upper half of the trajectory circle than in the lower half.
As seen in previous section, tracing a quarter of the trajectory circle is enough to extract physical parameters for astrometric microlensing such as $t_{\rm q}\; (=\beta t_{\rm E})$ and $\theta_{\rm E}/\beta$.
Thus, practically it is necessary to trace the image motion for the time interval $t_{\rm q}$, which was defined in section 2 as the time during which the image passes the upper quarter of the trajectory circle.
This duration $t_{\rm q}$ is given by (see figure 2)
\begin{equation}
t_{\rm q} = \beta t_{\rm E} = \frac{\beta_{\rm max} R_{\rm E}}{v_\perp} |\sin \phi|.
\end{equation}
Similarly to $\bar{t}_{\rm ast}$ (eq.[\ref{eq:t_ast_av_def}]), by performing the integration with respect to $\phi$ we can obtain the average of $t_{\rm q}$ as
\begin{equation}
\label{eq:t_q_av}
\bar{t}_{\rm q} = \frac{1}{2} \frac{\beta_{\rm max} R_{\rm E}}{v_\perp}.
\end{equation}
Comparing equations (\ref{eq:t_ast_av_with_beta}) and (\ref{eq:t_q_av}), one finds that $\bar{t}_{\rm q}= \bar{t}_{\rm ast}/\pi$, and thus a typical value of $\bar{t}_{\rm q}$ is $\sim$ 5 yr.
This means that one can obtain physical parameters of astrometric microlensing such as $\beta t_{\rm E}\; (=t_{\rm q})$ and $\theta_{\rm E} /\beta$ based on $\sim 5$ yr monitoring of the event,  which is rather reasonable for practical observations.

\subsection{Disk Microlensing}

Here we estimate the event duration for disk star lensing.
As a typical mass of disk stars, we take the mass of $M=0.3 M_\odot$, corresponding to a lower main-sequence star.
Since we mainly consider disk-star lensing at relatively high latitude (i.e., $b\ge 10^\circ$) in which the lens is likely to be a nearby star, we take the tangential velocity of 20 km s$^{-1}$, corresponding to a random velocity of disk star in the vicinity of the Sun.
With these parameters, the event duration for disk star microlensing is obtained as
\begin{eqnarray}
\bar{t}_{\rm ast} &=& 92 \left(\frac{M}{0.3 M_\odot}\right) \nonumber \\
 & & \times \left(\frac{v_\perp}{20 {\rm km\; s}^{-1}}\right)^{-1} \left(\frac{\theta_{\rm min}}{10 \mu{\rm as}}\right)^{-1}\; {\rm yr}.
\end{eqnarray}
Thus, the event duration for disk stars is much longer than that for halo MACHOs.
This applies for most of the sky regions except for the Galactic plane, and hence the discrimination of disk lensing from MACHO lensing is easy.
However, if one observes sources in the Galactic plane, the tangential velocity of the disk star is dominated by the galactic rotation, which is of $\sim 200$ km s$^{-1}$.
This indicates that in the galactic plane the event duration of disk lensing can be quite close to that of MACHO lensing, and hence sources in the middle of the galactic plane are not suitable for studying astrometric microlensing due to MACHOs,

%%%
\section{Optical Depth}
%%%
\subsection{Basic Equations}

Similarly to the optical depth of photometric microlensing (e.g., Paczynski 1986), the optical depth for the astrometric microlensing can be defined as follows (Miralda-Escude 1996),
\begin{equation}
\tau_{\rm ast} = \int \pi \beta_{\rm max}^2 R_{\rm E}^2\; \frac{\rho(D_{\rm d})}{M}\; dD_{\rm d}.
\end{equation}
Here $\rho$ is the density, $D_{\rm d}$ is the lens distance from the observer, and $M$ is the lens mass.
For simplicity here we assume that the lens mass $M$ is unique, as usually assumed in microlensing studies.

Substituting equations (\ref{eq:E-ring-approx}), (\ref{eq:angular-E-ring-approx}) and (\ref{eq:beta_max}), the optical depth (for distant sources) can be written as,
\begin{equation}
\label{eq:tau_ast}
\tau_{\rm ast} = \frac{16\pi G^2 M}{c^4 \theta_{\rm min}^2} \int \rho \; dD_{\rm d}.
\end{equation}
As is already known (e.g., Miralda-Escude 1996), the optical depth for astrometric microlensing is proportional to the lens mass $M$, in contrast to that for photometric lensing which is independent of the lens mass $M$.
It is also remarkable that the optical depth is simply proportional to the column density, $\int \rho\; dD_{\rm d}$.
This simple dependence on the column density comes from the approximation of distant source in equations (\ref{eq:E-ring-approx}) and (\ref{eq:angular-E-ring-approx}).
In more general case, the optical depth is proportional to $\int \rho (1-x^2)\; dD_{\rm d}$, where $x\equiv D_{\rm d}/D_{\rm s}$ is the fractional distance to the lens normalized with source distance (Dominik \& Sahu 2000).

\subsection{Halo Microlensing}

\begin{figure}
\vspace{5.8cm}
\epsfxsize=17pt
\epsfbox[30 50 80 100]{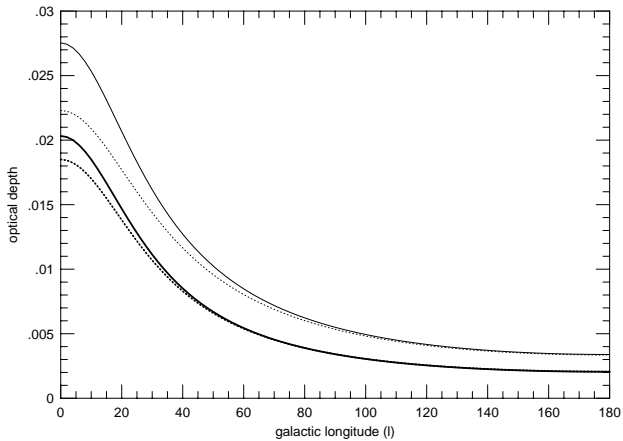}
\caption{Optical depth for astrometric microlensing due to MACHOs.
The mass of MACHO is assumed to be 0.5$M_\odot$.
Thick line corresponds to ($b$, $q$)=(0$^\circ$, 1), thin line to (0$^\circ$, 0.75), thick dashed line to (10$^\circ$, 1), and thin dashed line (10$^\circ$, 0.75).
The optical depth becomes larger with more flattened halo, because the halo density at the Galaxy's center increases with decreasing $q$.
}
\end{figure}

\begin{figure}
\vspace{5.8cm}
\epsfxsize=17pt
\epsfbox[30 50 80 100]{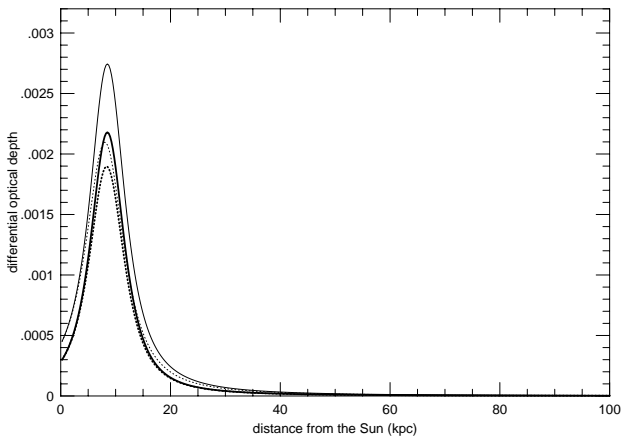}
\caption{Differential optical depth for astrometric microlensing due to MACHOs.
Notations are the same to those in figure 3.
}
\end{figure}

In order to calculate the optical depth due to MACHOs in the Galactic halo, here we introduce an axisymmetric logarithmic halo model (Binney \& Tremaine 1987) to describe the halo density distribution.
The density distribution of the axisymmetric logarithmic halo model is given by,
\begin{eqnarray}
\label{eq:halo-density}
\rho(R, z) &=& \frac{v_0^2}{4\pi G q^2} \nonumber \\
& & \!\!\!\!\! \times\, \frac{(2q^2+1)R_{\rm c}^2 + R^2 + (2-q^{-2})z^2}{(R_{\rm c}^2 + R^2 + z^2 q^{-2})^2}.
\end{eqnarray}
Here $R$ and $z$ describes the cylindrical coordinate, and $v_0$, $R_{\rm c}$, and $q$ are halo parameters which describe the circular rotation velocity, the core radius, and the halo flatness, respectively.
In the following analysis, we set $R_{\rm c}=5$ kpc and $v_0=200$ km s$^{-1}$, which are typical values used in microlensing studies.
We also introduce a cut-off radius beyond which the density vanishes.
In the present study the cut-off radius is set to be 100 kpc, but note that our result would be only slightly changed by varying the cut-off radius, as we will see later.
In addition to halo parameters in equation (\ref{eq:halo-density}), the mass of lens is necessary for optical depth calculation, as $\tau_{\rm ast}$ depends on the lens mass (equation [\ref{eq:tau_ast}]).
We set the MACHO mass $M=0.5M_\odot$, according to the recent result from the LMC microlensing (e.g., Alcock et al. 1997; 2000).
We also assume that the distance to the Galactic Center, $R_0$, is 8.5 kpc.

The optical depth for astrometric microlensing is calculated for two values of $q$; $q=1$ (spherically symmetric halo) and $0.75$ (flattened halo).
Figure 3 shows the optical depth variation with Galactic longitude $l$.
As seen in figure 3, the optical depth $\tau_{\rm ast}$ is quite high toward the galactic center, being $2\sim3 \times 10^{-2}$.
For a comparison, we note that the optical depth for photometric microlensing $\tau_{\rm ph}$ is $5.4 \times 10^{-6}$ for $l=0^\circ$ and $b=0^\circ$ ($q=1$ is assumed).
This high optical depth for astrometric microlensing can be easily understood from the fact that typical $\beta_{\rm max}$ is more than 70 (see previous section).
From the high value of $\tau_{\rm ast}$, one can expect that one of 30 to 50 distant sources seen around the galactic center is always being lensed, if the halo fully consists of MACHOs.
Recent results of LMC microlensing indicate that the halo MACHO fraction is around 25$\sim$50\%, and thus the optical depth may be 2 to 4 times smaller than what is obtained above.
However, the probability is still as high as $5\times 10^{-3}$, indicating that at least one of a few hundred sources is always being lensed.
Note that in figure 3 the optical depth is larger for flat halo case ($q=0.75$) than for spherical case ($q=1$).
This is because the halo density at the Galaxy's center increases with flattening of the halo.
For instance, at the Galaxy center ($R=0$, $z=0$) the halo density is proportional to $(2q^2+1)/q^2$, which is 3.778 for $q=0.75$ while 3 for $q=1$.

In order to study a typical lens distance, in figure 4 we plot the differential optical depth for astrometric microlensing which is given by
\begin{equation}
\frac{d\tau_{\rm ast}}{dD_{\rm d}}=\frac{16\pi G^2 M}{c^4 \theta_{\rm min}^2} \rho.
\end{equation}
Figure 4 shows that the largest contribution to the optical depth occurs at $D\approx 8.5$ kpc, i.e., around the Galactic center.
This is of course because that the halo density is largest at the Galactic center region.
As seen in figure 4, the differential optical depth beyond 50 kpc is quite small compared to the Galactic center region.
Therefore, the optical depth $\tau_{\rm ast}$ only slightly depends on the value of cut-off radius (which is assumed to be 100 kpc), provided it is sufficiently large.

\subsection{Disk Microlensing}

\begin{figure}[t]
\vspace{5.8cm}
\epsfxsize=17pt
\epsfbox[30 50 80 100]{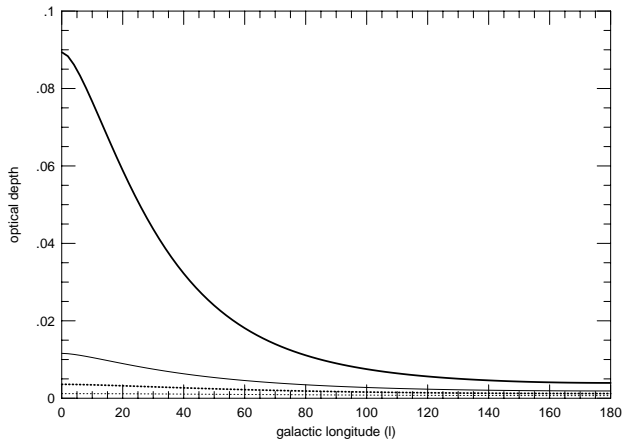}
\caption{Optical depth for astrometric microlensing due to disk stars.
The mass of disk stars is assumed to be 0.3$M_\odot$.
From the top to the bottom, four lines correspond to $b=0^\circ$, $5^\circ$, $10^\circ$, and $20^\circ$.
}
\end{figure}

\begin{figure}
\vspace{5.8cm}
\epsfxsize=17pt
\epsfbox[30 50 80 100]{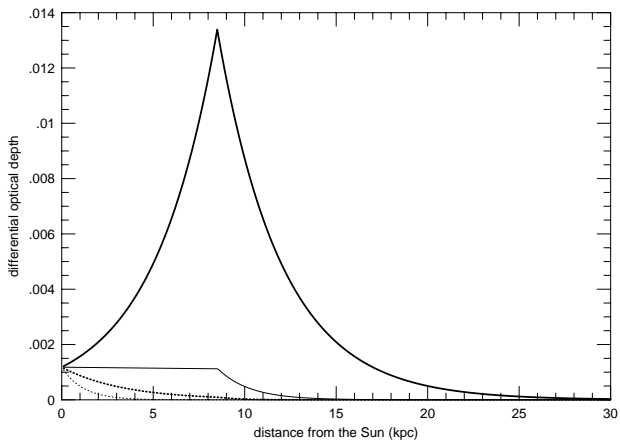}
\caption{Differential optical depth for astrometric microlensing due disk stars.
Notations are the same to those in figure 5.
}
\end{figure}

Since the stellar disk is a dominant component in the inner Galaxy, we have to take into account the contribution of stellar disk to astrometric microlensing.
In order to calculate the optical depth due to disk stars, here we assume a disk which has an exponential profile both in radial and vertical directions, namely,
\begin{equation}
\rho_{\rm d}(R,z) = \rho_0 \exp \left(-\frac{R-R_0}{d}-\frac{|z|}{h}\right).
\end{equation}
Here $\rho_0$ is the local disk density, and $d$ and $h$ are radial and vertical scale length, respectively.
We take typical values for these parameters as $\rho_0=0.08 M_\odot$, $d=3.5$ kpc,  and $h=300$ pc, which are the same to those used in previous studies (e.g., Hosokawa et al. 1997; Dominik \& Sahu 2000).
We show in figure 5 the optical depth for astrometric microlensing due to disk stars assuming $M=0.3M_\odot$.
As seen in figure 5, the optical depth is close to 0.1 at the galactic center, indicating that one of about 10 sources is always being lensed.
This optical depth by disk stars is consistent with previous studies such as Hosokawa et al.(1997).
If one observes a source relatively far from the galactic center but in the middle of the galactic plane ($b=0^\circ$), the optical depth due to disk star is larger than 0.02 for $l\le 50^\circ$ .
Therefore, as long as one observes sources in the Galactic plane, the optical depth is significantly higher than that for halo MACHOs, and hence sources in the galactic plane are not suitable for astrometric microlensing study of MACHOs (but of course they are useful for studying the galactic disk, particularly the stellar mass and the disk scale length; e.g., Hosokawa et al. 1993; Dominik \& Sahu 2000).
However, as seen in figure 5, the optical depth due to disk stars decreases quite rapidly with increasing the galactic longitude.
Thus a source at relatively high galactic latitude ($b\ge 10^\circ$) may be used to search for astrometric microlensing due to MACHOs.

Similarly to the halo case, in figure 6 we show the differential optical depth for disk stars.
As easily understood, the differential optical depth in the galactic plane ($b=0^\circ$) becomes maximum at the Galactic center.
On the other hand, for sources at relatively high galactic latitude, the differential optical depth for disk stars is largest at $D_{\rm d}=0$, and hence it is dominated by local stars located within a few kpc from the sun.
In such a case, one can discriminate the disk star microlensing from halo MACHO microlensing based on the event duration, as seen in the previous section.

\section{Event Rate}

The event rate is another important quantity in microlensing study, because the event rate is tightly related to the observational strategy of microlensing search.
Once the optical depth $\tau_{\rm ast}$ and the average event duration $\bar{t}_{\rm ast}$ are known, the event rate $\Gamma_{\rm ast}$ is obtained as,
\begin{equation}
\label{eq:gamma_ast}
\Gamma_{\rm ast} = \frac{\tau_{\rm ast}}{\bar{t}_{\rm ast}}.
\end{equation}
Substituting equations (\ref{eq:t_ast_av}) and (\ref{eq:tau_ast}) into above equation, the event rate for astrometric microlensing may be written as follows,
\begin{equation}
\Gamma_{\rm ast} = \frac{8 G v_\perp}{c^2 \theta_{\rm min}}\;\int \rho\; dD_{\rm d}.
\end{equation}
Remarkably, the event rate $\Gamma_{\rm ast}$ does not depend on the MACHO mass $M$, as is the case for the optical depth for photometric microlensing ($\tau_{\rm ph}$).
Also, the event rate $\Gamma_{\rm ast}$ is simply proportional to the column density $\int \rho\; dD_{\rm d}$.
These facts indicate that the event rate of astrometric microlensing is a good tracer of density distribution rather than the mass of individual lenses.

If we put into equation (\ref{eq:gamma_ast}) $\bar{t}_{\rm ast}=15$ yr and $\tau_{\rm ast}=1.5 \times 10^{-2}$, which corresponds to the optical depth at 20$^\circ$ away from the Galactic center, we obtain
\begin{eqnarray}
\Gamma_{\rm ast} &=& 1.0\times 10^{-3} \left(\frac{v_\perp}{200\; {\rm km\; s}^{-1}}\right) \nonumber \\
& & \quad\times \left(\frac{\theta_{\rm min}}{10\; \mu as}\right)^{-1} \; {\rm event\;\; yr}^{-1}.
\end{eqnarray}
Thus, if one observes 100 sources for 10 years, at least one event will be detected.
However, note that this calculation is for the case that the halo totally consists of MACHOs.
Since the halo MACHO fraction is estimated to be 25\%-50\% by recent studies (Alcock et al. 1997; 2000), one has to observe 2-4 times more sources or longer period to detect an event.
This indicates that it is necessary to monitor a few hundred sources for a few decades.
This number of source and the time span may seem to be long, but the frequency of monitoring can be fairly low because of the long event duration, and thus such an observation can be done with upcoming astrometric projects like VERA (we will discuss this in next section in detail).

For the disk star lensing, the event rate is expected to be much smaller than that for halo MACHO lensing because the event duration $t_{\rm ast}$ is significantly larger.
From figure 5, the optical depth for disk star lensing at $b=20^\circ$ is $1.2\times 10^{-3}$ (note that this value is almost independent of the galactic latitude $l$).
Using the event duration of 92 yr, one obtains an event rate for astrometric microlensing due to disk star as $1.3\times 10^{-5}$, which is smaller by two orders of magnitude than that for halo MACHO lensing.
Thus, the contribution of disk star lensing is negligible in the off-plane region.

\section{Implications for VERA}

In this section we discuss the implication of astrometric microlensing for VERA.
As we have briefly mentioned in section 1, VERA is a new VLBI array to measure the relative position of two adjacent sources based on the phase-referencing technique.
Although VERA aims at astrometry with 10 $\mu$as-level accuracy similarly to other space missions (e.g., SIM, GAIA), VERA is very unique as it is an array of ground-based radio telescopes.
The advantage of using VERA for astrometric search is that the project life time is much longer than that of other space missions, being $\sim$20 years (as for space missions like SIM and GAIA, the mission lifetime is around 5 years).
As we have seen in the previous sections, the event duration is as long as 10 years, and thus a monitoring span exceeding 10 years is indispensable for studying astrometric microlensing event.

As we have shown previously, to detect an astrometric microlensing event, a few hundred sources should be monitored for a few decades.
If one wants to detect more events for statistical studies, the number of sources to be observed becomes larger.
Thus, it is important to examine if there are enough number of sources for astrometric microlensing search with VERA.
The main target of VERA are compact continuum sources and maser sources.
The compact continuum sources are mainly distant radio galaxies and QSOs, and thus they are suitable to astrometric microlensing studies.
On the other hand, the maser sources are mostly late-type stars and star forming regions in the Galaxy, and so unsuitable for the purpose of the present study.
As for continuum sources, about 2000 VLBI sources (which are compact enough to observe with VLBI) are currently known (e.g., ICRF catalog, Ma et al. 1998; VLBA Calibrator Survey, Peck \& Beasley 1998).
At high galactic latitude where VLBI sources are searched relatively deeply, the average number of VLBI sources is close to 0.08 per square degree.
This number roughly corresponds to one VLBI source in a circle with radius of 2$^\circ$.
If the same source distribution applies to the sky region toward the galactic center, one can expect to find about 100 sources within 20$^\circ$ of the Galactic center, and 220 sources within 30$^\circ$ of the Galactic center, indicating that there are enough number of VLBI sources that can be observed with VERA.
Moreover, there are several hundreds of VLBI source candidates like those in Jodrel-Bank VLA Astrometric Survey (JVAS; Patnaik et al. 1992; Browne et al. 1998; Wilkinson et al. 1998).
By a deeper and more massive survey, one can expect to increase the number of VLBI sources that can be used for astrometric microlensing search.
Hence, one can expect to have a plenty of VLBI sources around the galactic center regions.

Next we estimate the time required for such a monitoring observation.
The signal-to-noise ration (S/N) of VLBI can be estimated using the following equation (e.g., Thompson et al 1986),
\begin{equation}
R_{\rm S/N} = \frac{\eta}{2k}\sqrt{\frac{A_1 A_2}{T_{\rm sys1}T_{\rm sys2}}}\; \sqrt{2\Delta\nu\; t_{\rm i}}\; S.
\end{equation}
Here $A$ is the effective antenna collecting area, $T_{\rm sys}$ is the system noise temperature, $\Delta\nu$ is the bandwidth, $t_{\rm i}$ is the integration time, $k$ is the Boltzmann constant, $\eta$ is the quantization efficiency, and $S$ is the source flux (subscripts denote the observational stations).
The major specifications of VERA telescopes are currently anticipated as follows; the effective collecting area $A$ of 157 m$^2$ (assuming 20m diameter and antenna efficiency of 0.5), system noise temperature $T_{\rm sys}$ of 120 K, bandwidth $\Delta\nu$ of 256 MHz, and the quantization efficiency $\eta$ of 0.88 (two-bit quantization).
With these values, the S/N ratio $R_{\rm S/N}$ is $\sim$ 16 for 100 mJy source with integration time  $t_{\rm i}$ of 300 seconds.
Since the accuracy of fringe phase measurement is determined as $1/R_{\rm S/N}$ in radian, its accuracy will be 0.063 radian, or 3.6 degree.
This corresponds to an error in geometric delay of 0.13 mm ($=3.6\times \pi /180 \times \lambda$, where $\lambda$ is the wavelength) at the wavelength of 1.3 cm, which is one of the major frequency band of VERA.
The position accuracy can be roughly estimated as (0.13 mm/2300 km)=5.65$\times 10^{-11}$ radian, which equals to 11.7 $\mu$as (2300 km is the longest baseline of VERA).
Thus, for sources with a flux of $\sim$ 100 mJy, the thermal error can be reduced to 10 $\mu$as level within 5 minutes.

If one spends 20 minutes for a single measurement (one measurement of separation of two adjacent sources), the total number of sources that can be observed in 24 hours exceeds 100 (note that two sources are observed in a single measurement).
Thus, 2-3 days are enough for observing a few hundred sources.
In practice, one separation measurement for each pair is not enough for detecting a position shift of $\sim$10 $\mu$as.
To achieve higher position accuracy (i.e., a few $\mu$as), we may need 10 sets of observations of a few hundred sources, which requires total observation of 20-30 days.
On the other hand, typical timescale for a image to move along a quarter of the trajectory is about 5 years, and so the interval of such observations can be as long as six months.
If we spend 30 days for astrometric microlensing in six months, this requires 15\% of the telescope time to be dedicated to such a monitoring observation.
This is indeed massive expense, but is worth spending when one considers the significance of scientific outputs; in addition to astrometric microlensing, one can establish a radio reference frame with an accuracy of 10 $\mu$as, which is more accurate by two orders of magnitude than the reference frame currently achieved, and also one can measure proper motions of nearby AGNs (up to a few tens Mpc), from which 3 dimensional structure of the local universe can be studied.

One possible way to incorporate such a massive microlensing search with VERA is to perform a geodetic observation based on the dual-beam observation.
The geodetic observation is necessary to determine the baseline parameters with sufficient accuracy.
Usually, a geodetic observation is made at S (2GHz) and X (8GHz) bands with a single beam.
In case of S/X geodetic observation, the baseline parameters and absolute position of radio sources with an accuracy of 1 mas are solved based on a few hundred observations of radio sources.
On the other hand, baseline measurements are also possible based on the relative fringe phase of two sources that are observed with dual beam system of VERA.
In case of dual-beam geodetic observation, what can be obtained are the baseline parameters and {`}relative{'} position of adjacent two sources with an accuracy of 10 $\mu$as level.
Thus, if dual-beam geodetic observation is performed, one can obtain relative position of radio sources for a few hundred sources with an accuracy of 10 $\mu$as.
Currently, VERA is planned to perform one-day geodetic observation every week, and thus the frequency of the observation is just the same to what we need for astrometric microlensing search (i.e., 15\% of the telescope time).
Therefore, dual-beam geodetic observation, if possible, is more practical for the massive monitoring of astrometric microlensing.

\section{Conclusion}

We have investigated the properties of astrometric microlensing of distant sources due to MACHOs.
We have shown that in case of astrometric microlensing of distant sources, the event duration depends only on two unknown parameters, the lens mass $M$ and the tangential velocity $v_\perp$, in contrast to that for photometric microlensing of nearby stars.
We have also shown that a typical event duration of astrometric microlensing is 15 years, and also that the event rate of astrometric microlensing is larger than 2.5$\times 10^{-4}$ within $20^\circ$ of the Galactic center, assuming that 25\% of Galaxy's halo is made up with MACHOs.
This implies that one can detect an astrometric microlensing event if a few hundred sources are monitored for 20 years.
Detections of astrometric microlensing will be possible with VERA within decades, and thus astrometric microlensing will become a new tool to study the nature of MACHOs.

%\clearpage
\section*{References}
\def\re{\hangindent=1pc \noindent}

\re Alcock C. et al. 1993, Nature 365, 621

\re Alcock C. et al. 1997, ApJ 486, 697

\re Alcock C. et al. 2000, ApJ 542, 281

\re Aubourg E. et al. 1993, Nature 365, 623

\re Boden A. F., Shao M., Van BurenD. 1998, ApJ 502, 538

\re Binney J. Tremaine S. 1987, {\it Galactic Dynamics} (Princeton University Press, Princeton)

\re Browne I.W.A., Patnaik A.R., Wilkinson P.N., Wrobel J.M. 1998, MNRAS 293, 257

\re Dominik M., Sahu K. 2000, ApJ 534, 213

\re Evans N. W., Gyuk G., Turner M. S., Binney J. 1998, ApJ 501, L45

\re Hog E., Novikov, I. D., Polnarev A. G. 1995, A\&A 294, 287

\re Honma M, Kan-ya Y. 1998, ApJ 503, L139

\re Honma M., Kawaguchi N., Sasao T. 2000, in Proc. SPIE Vol.4015 Radio Telescope, ed H. R. Buthcer, p624 - p631

\re Hosokawa M., Ohnishi K., Fukushima T. Takeuti M. 1993, A\&A 278, L27

\re Hosokawa M., Ohnishi K., Fukushima T. 1997, AJ 114, 1508

\re Kawaguchi N., Sasao T., Manabe S. 2000, in Proc. SPIE Vol.4015 Radio Telescope, ed H. R. Buthcer, p544 - p551

\re Ma C., Arias E.F., Eubanks T.M., Fey A.L., Gontier A.M., Jacobs C.S., Sovers O.J., Archinal B.A., Charlot P. 1998, AJ 116, 516 (ICRF catalog)

\re Miralda-Escude J., 1996, ApJ 470, L113

\re Miyamoto M., Yoshii Y. 1995, AJ 110, 1427

\re Paczynski B. 1986, ApJ 304, 1

\re Paczynski B. 1996, Acta Astronomoica 46, 291

\re Paczynski B. 1998, ApJL 494, L23

\re Patnaik A.R., Browne I.W.A., Wilkinson P.N., Wrobel J.M. 1992, MNRAS 254, 655

\re Peck A.B., Beasley A.J. 1998, in Radio Emission from Galactic and Extragalactic Compact Sources, IAU Colloquium 464, ed, J.A. Zensus, G.B. Taylor, J.M. Wrobel, ASP Conf. Ser. 144, p155 (VLBA calibrator survey)

\re Sahu K. 1994, Nature 370, 275

\re Sasao T. 1996, in proceedings of 4th APT Workshop, ed. E. A. King, p94 - p 104

\re Thompson A. R., Moran J. M., Swenson G. W. Jr. 1998, Interferometry and Synthesis in Radio Astronomy (Krieger publishing company, Florida)
 
\re Walker 1995 M. A. 1995, ApJ 453, 37

\re Wilkinson P.N., Browne I.W.A., Patnaik A.R., Wrobel J.M., Sorathia B. 1998, MNRAS 300, 790

\re Zaritsky, D., Lin D. N. C. 1997, AJ 114, 2545

\re Zhao H. 1998, MNRAS 294, 139

\end{document}